\documentclass[twoside,12pt]{article}
\usepackage{amssymb,graphicx,macros2e}
\usepackage{times}

\skip\footins 39pt plus 4pt minus 2pt
\def\footnoterule{\kern-19pt\hrule width.5in\kern18.6pt}%
\def\rmp#1#2#3{Rev. Mod. Phys. {\bf #1}, #2 (#3)}

\def\ptpo#1#2#3{Prog. Theor. Phys. {\bf #1}, #2 (#3)}

\def\beq{\begin{equation}}
\def\eeq{\end{equation}}
\def\bea{\begin{eqnarray}}
\def\eea{\end{eqnarray}}
\def\bc{\begin{center}}
\def\ec{\end{center}}
\def\rme{\mathrm{e}}
\newcommand{\rmd}{\mathrm{d}}

\newif\ifpdf
	\ifx\pdfoutput\undefined
	\pdffalse 
	\newcommand{\fig}[2]{\includegraphics[width=#1]{./figures/#2.eps}}
	
        \newlength{\bilderlength}

\else
	\pdfoutput=1 
	\newcommand{\fig}[2]{\includegraphics[width=#1]{./figures/#2.pdf}}
	
	\newlength{\bilderlength}

	\pdftrue
\fi

\begin{document}

\title{\vspace{-1.5cm}\sffamily\Large\bfseries Perturbative
Linearization of Reaction-Diffusion Equations}

\author{{\sffamily\bfseries Sanjay Puri$^1$ and Kay J\"org Wiese$^2$} 
\medskip \\
\small $^1$School of Physical Sciences, Jawaharlal Nehru University,
New Delhi -- 110067, India. \\
\small $^2$ KITP, Kohn Hall, University of California at Santa Barbara, Santa
Barbara, CA, 93106-4030, U.S.A.}
\date{\small\today}
\maketitle
\begin{abstract}
We develop perturbative expansions to obtain solutions for
the initial-value problems of two important reaction-diffusion systems,
viz., the Fisher equation and the time-dependent Ginzburg-Landau (TDGL)
equation. The starting point of our expansion is the corresponding
singular-perturbation solution. This approach transforms the solution
of nonlinear reaction-diffusion equations into the solution of a 
hierarchy of linear equations. Our numerical results
demonstrate that this hierarchy rapidly converges to the exact solution.
\end{abstract}

\section{Introduction}

Many physical problems are described by nonlinear partial differential
equations (pdes). In particular, much research interest has focused
upon two classes of nonlinear pdes, viz., \\
(i) {\it soliton-bearing equations}, which arise in the context of
completely-integrable infinite-dimensional Hamiltonian systems
\cite{as,pur}; and \\
(ii) {\it reaction-diffusion equations}, which arise in the context
of pattern-forming systems where local reactions are combined
with spatial diffusion \cite{mei,kur,saa,ch}.

There is no general framework for obtaining the
solution of the initial-value problem
for an arbitrary nonlinear pde. For the class of soliton-bearing equations
in (i), powerful techniques like the {\it inverse scattering transform} and
{\it Backlund transformations} enable the solution of these equations for
arbitrary initial conditions \cite{as,pur}. Essentially, these
methods reduce the problem of solution of a nonlinear soliton equation
to a sequence of linear equations. In a related context, a classic
example of linearization is provided by the Cole-Hopf transformation
\cite{hc}, which transforms the nonlinear Burgers' equation \cite{bur}
into the linear diffusion equation.

For the class of reaction-diffusion equations in (ii) 
above, there are as yet no systematic methods of linearization.
These pdes have the general form:
\begin{equation}
\label{rd}
\partial_{t} \psi (\vec r,t) = f (\psi) + \nabla^2 \psi ,
\end{equation}
where $\psi (\vec r,t)$ is an order-parameter field, e.g., population
density, chemical concentration, magnetization, which depends
on space $(\vec r)$ and time $(t)$. The order parameter may be either
scalar or vector, depending upon the number of variables which describe
the physical system. The order parameter evolves 
in time due to a local reaction, described by the nonlinear term 
$f(\psi)$, in conjunction with spatial diffusion. Reaction-diffusion
equations are ubiquitous in pattern-forming systems,
ranging from chemical and biological physics to materials science and
metallurgy \cite{mei,kur,saa,ch}.

An important example of Eq.~(\ref{rd}) is the Fisher equation with
$f(\psi) = \psi-\psi^2$, which describes the growth and saturation of
a species population \cite{fis,kpp}. Another important reaction-diffusion
model is the time-dependent Ginzburg-Landau (TDGL) equation with
$f (\vec \psi) = \vec \psi - |\vec \psi|^2 \vec \psi$ \cite{hh,bra}, where
$\vec \psi$ is an $n$-component vector, $\vec \psi
\equiv (\psi_1, \psi_2,..., \psi_n)$. The case with $n=1$ describes
the phase ordering dynamics of a 
ferromagnet which has been suddenly quenched from the
paramagnetic phase to the ferromagnetic phase. The TDGL equation with
$n=2$ describes phase
ordering dynamics in superconductors, superfluids and liquid crystals
\cite{bra}. In this paper, we will investigate the perturbative
linearization of the Fisher and TDGL equations for arbitrary
initial conditions.

This paper is organized as follows. Section \ref{Overview of
Analytical Results} provides an overview of relevant analytical
results, primarily in the context of the Fisher equation. Section
\ref{Analytical and Numerical Results} discusses our linearization
scheme and presents detailed numerical results therefrom. Finally,
Section \ref{Summary and Discussion} concludes this paper with a
summary and discussion of our results.

\section{Overview of Analytical Results}
\label{Overview of Analytical Results}
Let us first consider the Fisher equation in an infinite domain, which has
the following form \cite{fis}:
\begin{equation}
\label{fe}
\partial_{t} \psi (\vec r,t) = \psi - \psi ^2 + \nabla^2 \psi .
\end{equation}
In general, Eq.~(\ref{fe}) is supplemented with some arbitrary 
initial condition $\psi (\vec r,0)$. Typically, we are interested in 
the case with $\psi \geq 0$, as the order parameter describes population 
density which cannot be negative. As a matter of fact, 
Eq.~(\ref{fe}) is unstable for $\psi < 0$.
The homogeneous solution $\psi^*=0$ is an unstable fixed point (FP) of
the dynamics. Fluctuations about $\psi^*=0$ diverge exponentially and 
saturate to the stable FP, $\psi^*=1$.

There is no general solution available for the initial-value problem
of Eq.~(\ref{fe}). However, some important analytical results are 
known for the case with dimensionality $d=1$. Kolmogorov {\it et al.} 
\cite{kpp} found that the Fisher equation has stable traveling-wave 
solutions $\psi (x,t) \equiv \psi (x-vt)$ (called {\it clines}),
which are domain walls with \\
(a) velocity $v \geq 2$, and $\psi (-\infty,t)=1$, $\psi (\infty,t)=0$; or \\
(b) velocity $v \leq -2$, and $\psi (-\infty,t)=0$, $\psi (\infty,t)=1$. \\
The qualitative forms of these solutions are easily obtained
through a phase-portrait analysis, but the explicit analytic forms
are unknown.

In more general work, Aronson and Weinberger \cite{aw} considered the $d=1$
version of Eq.~(\ref{rd}). They focused on functions $f(\psi)$
which satisfy the conditions
\begin{eqnarray}
f (\psi) & \geq & 0 ~~~~ {\hbox{for}} ~~~~ \psi \in [0,1], 
\nonumber\\
f (0) & = & f (1) = 0, \nonumber\\
f'(0) & > & 0 > f'(1).
\end{eqnarray}
These authors demonstrated that a broad class of initial conditions
$\psi (x,0)$, with sharp interfaces, converge to a traveling-wave 
solution with a definite speed $c^*$, which satisfies
\bea
2[f'(0)]^{1/2} & \leq & c^* \leq 2 L^{1/2}, \nonumber \\
L & = & {\hbox{sup}} \left[ {f(\psi) \over \psi} \right],~~~~ \psi \in [0,1].
\eea
A physical explanation of this velocity-selection principle
has been formulated by various authors \cite{msh}, and is 
often referred to as the {\it marginal
stability hypothesis}. However, this approach has proven inadequate
in various applications. A more comprehensive understanding, based on
structural-stability arguments, is due to Paquette {\it et al.} \cite{pcgo}.

In the case of the Fisher equation, the Aronson-Weinberger result 
demonstrates that a large class of initial conditions
converge to the cline solution with $v=\pm 2$, whose functional form is
unknown as yet. An analytic form for a cline solution was first
obtained by Ablowitz and Zeppetella \cite{az} for $v = 5/\sqrt{6} >2$.

It is of obvious interest to obtain analytic forms for other 
cline solutions, particularly the case $v = \pm 2$. It is of 
even greater interest to obtain a general
solution for the initial-value problem of Eq.~(\ref{fe}). In this context,
Puri {\it et al.} \cite{ped} have used singular-perturbation techniques,
developed by Suzuki \cite{suz} and Kawasaki {\it et al.} (KYG) \cite{kyg}
in the context of the TDGL equation, to obtain an approximate solution
for the initial-value problem of the Fisher equation:
\bea
\label{sp0}
\tilde {\psi_0} (\vec r,t) & = & {\psi_L (\vec r,t) 
\over 1+\psi_L (\vec r,t)}, \nonumber \\
\psi_L (\vec r,t) & = & \rme^{t(1+\nabla^2)} \psi (\vec r,0) ,
\eea
where $\psi_L (\vec r,t)$ is the solution of the linear part of the Fisher
equation, and diverges with time.

The approximate solution in Eq.~(\ref{sp0}) has a number of attractive
features. For example, it is obtained 
for arbitrary dimensionality. Furthermore,
initial conditions with sharp interfaces 
evolve into a traveling-wave front with
asymptotic speed $v=2$, in accordance with the Aronson-Weinberger result.
The approach to the asymptotic velocity is as follows \cite{ped}:
\begin{equation}
v(t) \simeq 2 - {d \over 2t}.
\end{equation}
Unfortunately, this is not in agreement with the exact result for the $d=1$
Fisher equation obtained by Bramson \cite{bram}:
\begin{equation}
v(t) \simeq 2 - {3 \over 2t}.
\end{equation}
More generally, the fronts obtained using the singular-perturbation
approximation are appreciably sharper than the exact result (obtained
numerically) \cite{ped}. Furthermore, the solution in Eq.~(\ref{sp0})
is unable to accurately resolve front-front interactions \cite{ped}.
An improved approximation has been proposed by Puri and Bray \cite{pb}.
However, this is only valid for the less interesting class of initial
conditions where the order parameter $\psi > 0$ everywhere, i.e., the
evolving system is not characterized by the formation and
interaction of fronts.

Finally, it is also relevant to discuss the singular-perturbation 
solution for the TDGL equation:
\begin{equation}
\label{gl}
\partial_{t} \psi (\vec r,t) = \psi - \psi ^3 + \nabla^2 \psi .
\end{equation}
As mentioned earlier, the corresponding solution was obtained by
KYG \cite{kyg}, who generalized a diagrammatic technique developed by
Suzuki \cite{suz}. The approximate solution for the scalar TDGL 
equation is as follows:
\begin{eqnarray}
\label{glsp0}
\tilde {\psi_0} (\vec r,t) & = & {\psi_L (\vec r,t) 
\over \sqrt{1+\psi_L (\vec r,t)^2}}, \nonumber \\
\psi_L (\vec r,t) & = & \rme^{t(1+\nabla^2)} \psi (\vec r,0) .
\end{eqnarray}
Puri and Roland \cite{pr} have obtained the singular-perturbation
solution for the 2-component TDGL equation. This result has been
generalized by Bray and Puri \cite{bp} and Puri \cite{puri} to obtain
the time-dependent structure factor for the $n$-component TDGL
equation. As in the case of the Fisher equation, the approximate
solution in Eq.~(\ref{glsp0}) is characterized by fronts which are
too sharp. Furthermore, the singular-perturbation solution is
unable to properly resolve domain-wall interactions \cite{ped}.

In this paper, we undertake a perturbative improvement of the 
singular-perturbation solution of the Fisher and TDGL equations. 
We develop a hierarchy of linear equations which rapidly converges
to the exact solution. Our primary goal is methodological, viz., 
demonstrating the equivalence between nonlinear reaction-diffusion
equations and a sequence of linear equations.

\section{Analytical and Numerical Results}
\label{Analytical and Numerical Results}
\subsection{Fisher Equation}
The solution of the Fisher-equation with constant initial conditions
suggests the  following nonlinear transformation:
\begin{equation}
\label{nt}
\psi (\vec r,t) = {\phi (\vec r,t) \over 1+\phi (\vec r,t)},
\end{equation}
where $\phi > 0$ is an auxiliary field \cite{bra}. We confine
ourselves to the physically interesting case with $1 > \psi \geq
0$. The corresponding pde satisfied by $\phi (\vec r,t)$ is 
\bea
\label{phi} \partial_{t}\phi (\vec r,t) & = & \phi + \nabla^2 \phi -
{2 (\nabla \phi)^2 \over 1+\phi}, \nonumber \\
\phi (\vec r,0) & = & {\psi (\vec r,0) \over 1-\psi (\vec r,0)}.
\eea
The singular-perturbation approximation is equivalent to 
dropping the nonlinear
term on the right-hand-side (RHS) of Eq.~(\ref{phi}). In that case, the
solution (modified slightly from Eq.~(\ref{sp0})) is
\bea
\label{sp}
\psi_0 (\vec r,t) & = & {\phi_0 (\vec r,t) \over 1+\phi_0
(\vec r,t)} , \nonumber \\ 
\phi_0 (\vec r,t) & = & \rme^{t(1+\nabla^2)}
\left[ {\psi (\vec r,0) \over 1-\psi (\vec r,0)} \right] .
\eea
The approximate solution in Eq.~(\ref{sp}) reduces to that in Eq.~(\ref{sp0})
in the limit of small $\psi (\vec r,0)$. However, in contrast to the earlier
solution, Eq.~(\ref{sp}) yields the exact solution in the homogeneous case
$\psi (\vec r,0) = \bar \psi (0)$, viz.,
\begin{equation}
\bar {\psi_0} (t) = {\rme^t \bar \psi (0) \over 1 - 
\bar \psi (0) + \rme^t \bar \psi (0)}.
\end{equation}
We will use the approximate solution in 
Eq.~(\ref{sp}) as the starting point of a
perturbative expansion, which yields a hierarchy of linear equations.
First, notice that the exact pde obeyed by the solution 
$\psi_0 (\vec r,t)$ is
\begin{equation}
\partial_{t} \psi_{0} (\vec r,t) = \psi_0 - \psi^2_0 + \nabla^2 \psi_0 + 
{2 (\nabla \psi_0)^2 \over 1-\psi_0} .
\end{equation}
We decompose the exact solution of the Fisher 
equation as $\psi = \psi_0+\theta_0$,
where $\theta_0$ is the correction to the singular-perturbation solution.
Then, the pde obeyed by $\theta_0 (\vec r,t)$ is
\begin{equation}
\label{theta}
\partial_{t}{\theta_0} (\vec r, t) = (1-2\psi_0) \theta_0 - \theta_0^2 + 
\nabla^2 \theta_0 - {2 (\nabla \phi_0)^2 \over (1+\phi_0)^3}.
\end{equation}
We assume that $\theta_0$ is small, and break 
it up as $\theta_0=\psi_1+\theta_1$,
where $\psi_1$ solves the linear part of Eq.~(\ref{theta}):
\begin{equation}
\label{psi1}
\partial_{t}{\psi_1} (\vec r, t) = (1-2\psi_0) \psi_1 + \nabla^2 \psi_1 - 
{2 (\nabla \phi_0)^2 \over (1+\phi_0)^3} .
\end{equation}
The exact pde obeyed by $\theta_1 (\vec r,t)$ is then
\begin{equation}
\partial_{t}{\theta_1} (\vec r, t) = [1-2(\psi_0 + \psi_1)] \theta_1 - 
\theta^2_1 + \nabla^2 \theta_1 - \psi^2_1,
\end{equation}
which has the same general form as Eq.~(\ref{theta}).

This process can be continued indefinitely. At the 
$n^{\mathrm{th}}$ level, we decompose as
$\theta_{n-1}=\psi_n+\theta_n$, where $\psi_n$ solves the linear version of
the pde obeyed by $\theta_{n-1}$. The corresponding pde obeyed by $\theta_n$
is then
\begin{equation}
\partial_{t} {\theta_n} (\vec r, t) = \left( 1-2 \sum\limits^n_{k=0} 
\psi_k \right) \theta_n - \theta^2_n + \nabla^2 \theta_n - \psi^2_n.
\end{equation}
The approximate perturbative solution at the $n^{\mathrm{th}}$ 
level of iteration is then obtained as
\begin{equation}
\label{psol}
\psi (\vec r, t) \simeq \sum\limits^n_{k=0} \psi_k (\vec r, t) .
\end{equation}
The general form of the pdes obeyed by $\psi_k$ (for $k \geq 1$) is
\begin{equation}
\label{lin}
\partial_{t}{\psi_k} (\vec r, t) = a(\vec r, t) \psi_k + \nabla^2 \psi_k + 
b(\vec r, t) ,
\end{equation}
where $a(\vec r, t)$ and $b(\vec r, t)$ are functions of space
and time. A formal solution of Eq.~(\ref{lin}) reads (somewhat symbolically):
\begin{equation}
\label{form} \psi_k  = \Big(\partial_{t}-\nabla^{2}-a 
\Big)^{\!-1}\, b \ .
\end{equation}
It can be expanded in a Taylor-series in $a$. Denoting $R (\vec x,t)$ 
as the functional inverse of $(\partial_{t}-\nabla^{2})$, such that $
(\partial_{t}-\nabla^{2}) R (\vec x,t) = \delta (\vec x)\delta (t)$,
Eq.~(\ref{form}) can be written as
\begin{eqnarray}
 \psi_k (\vec r,t) &=& \int\limits_{0}^{t}\rmd t'\int \mathrm{d}
\vec r\,' R (\vec r-\vec r\,',t-t') b (\vec r\,',t')\nonumber \\
&&+\int\limits_{0}^{t}\rmd t'\int \mathrm{d}
\vec r\,'\int\limits_{0}^{t'}\rmd t''\int \mathrm{d}
\vec r\,''  R (\vec r-{\vec
r}\,',t-t') a (\vec r\,',t') R (\vec r\,'-\vec r\,'',t'-t'') b ({\vec
r}\,'',t'') + \dots\nonumber \\
\end{eqnarray}
Note that for $b (\vec r,t)=\delta (t)~\phi (\vec r,0)$ and $a=1$, we
recover the solution to the linear part of Eq. (\ref{phi}). 

We will demonstrate shortly that this perturbative expansion converges 
very rapidly to the exact solution. However, we stress that our interest
in such an expansion is more methodological than operational, i.e.,
the above procedure converts the problem of solution of
the nonlinear Fisher equation to a hierarchy of linear equations. Of course,
the same procedure would serve to improve any approximate solution -- the
singular-perturbation result is merely a convenient starting point.

\subsection{Time-Dependent Ginzburg-Landau Equation}
Before we present our numerical results, it is relevant to discuss the
corresponding procedure for the scalar TDGL equation. (The
generalization to the $n$-component case is relatively
straightforward.) The corresponding nonlinear transformation is
obtained by solving the TDGL-equation for constant initial conditions
and reads  
\beq \psi (\vec r, t) = {\phi (\vec r, t) \over
\sqrt{1+\phi (\vec r, t)^2}}\ , 
\eeq 
where $\phi$ is the appropriate auxiliary field. In this case, we
confine ourselves to the physically interesting case with $| \psi | <
1$.  The pde satisfied by $\phi (\vec r, t)$ is \cite{bra}
\begin{equation}
\label{glphi}
\partial_{t} \phi (\vec r, t) = \phi + \nabla^2 \phi - 
{3\phi (\nabla \phi)^2 \over 1+\phi^2}.
\end{equation}
As before, the singular-perturbation solution is obtained by 
neglecting the nonlinear term on the RHS of Eq.~(\ref{glphi}) as follows:
\begin{eqnarray}
\label{glsp}
\psi_0 (\vec r, t) & = & {\phi_0 (\vec r, t) \over 
\sqrt{1+\phi_0 (\vec r, t)^2 }} ,
\nonumber \\
\phi_0 (\vec r, t) & = & \rme^{t(1+\nabla^2)}
\left[ {\psi (\vec r,0) \over \sqrt{1-\psi (\vec r,0)^2}} \right] .
\end{eqnarray}
This solution constitutes an improvement over the solution in 
Eq.~(\ref{glsp0}) as it is exact in the homogeneous case. 
The exact pde satisfied by $\psi_0$ is then
\begin{equation}
\partial_{t}{\psi_0} (\vec r, t) = \psi_0 - \psi^3_0 + \nabla^2 \psi_0 + 
{3\psi_0 ( \nabla \psi_0)^2 \over 1-\psi^2_0}.
\end{equation}
We decompose the solution of the TDGL equation as $\psi=\psi_0+\theta_0$,
where the pde obeyed by $\theta_0 (\vec r, t)$ is
\begin{equation}
\partial_{t}{\theta_0} (\vec r, t) = (1-3\psi^2_0) \theta_0 - 
3\psi_0 \theta_0^2 - \theta_0^3 + \nabla^2 \theta_0
- {3\psi_0 ( \nabla \psi_0)^2 \over 1-\psi^2_0}.
\end{equation}
As before, we designate the solution of the linear part of this equation
as $\psi_1 (\vec r, t)$. If we decompose the overall solution
as $\theta_0=\psi_1+\theta_1$, the pde obeyed by $\theta_1 (\vec r, t)$ is
\begin{equation}
\partial_{t}{\theta_1} (\vec r, t) = [1-3 (\psi_0 + \psi_1)^2] \theta_1 - 
3(\psi_0 + \psi_1) \theta^2_1
- \theta^3_1 + \nabla^2 \theta_1 - (3\psi_0 + \psi_1) \psi_1^2 .
\end{equation}
Again, we can develop an infinite hierarchy of equations. The pde for 
$\theta_n (\vec r, t)$ at the $n^{\mathrm{th}}$ level of
this hierarchy is
\begin{equation}
\partial_{t}{\theta_n} (\vec r, t) = \left[ 1-3 \left( \sum\limits^n_{k=0} 
\psi_k \right)^2 \right]
\theta_n - 3\left( \sum\limits^n_{k=0} \psi_k \right) 
\theta^2_n - \theta^3_n +
\nabla^2 \theta_n - \left( 3 \sum\limits^{n-1}_{k=0} \psi_k
+ \psi_n \right) \psi^2_n .
\end{equation}
Therefore, the problem of solution of the nonlinear 
TDGL equation is again reducible to the solution of an infinite 
hierarchy of linear equations for $\psi _k (\vec r, t)$. 
The approximate perturbative solution
at the $n^{\mathrm{th}}$ level is given by Eq.~(\ref{psol}).

\subsection{Numerical Results for the Fisher Equation}

In this subsection, we will numerically examine the convergence properties
of the hierarchy of linear equations presented above. 
For the sake of brevity, we
confine ourselves to presenting numerical results for the Fisher equation
in $d=1,2$. Similar results are obtained for the TDGL equation also.

Our numerical results for the Fisher equation, referred to as the ``exact
solution'' subsequently, were obtained by implementing an Euler-discretized
version of Eq.~(\ref{fe}) (with an isotropic Laplacian) in $d=1,2$. In both
cases, we used periodic boundary conditions. The discretization mesh sizes 
in $d=1$ were $\Delta t=0.001$ and $\Delta x=0.1$ in time and space, 
respectively. The corresponding mesh sizes in $d=2$ were 
$\Delta t=0.001$ and $\Delta x=0.2$. The lattice size was
$N_1 = 40000$ in $d=1$, and $N_2^2 = 1000^2$ in $d=2$. The spatial coordinates
are $x \in [-2000,2000]$ in $d=1$; and $x,y \in [-100,100]$ in $d=2$.
Our perturbative solutions for
$\psi_0, \psi_1, \psi_2$, etc. were also obtained numerically by solving the
relevant linearized equations. The discretization meshes and lattice sizes
for the perturbative solutions are identical to those described above.

Figure 1(a) shows the profile of a front arising from a seed initial
condition ($\psi (x,0) = 0.05 \delta (x)$) for the $d=1$ Fisher 
equation. The evolution gives two equivalent clines moving in
opposite directions -- we focus on the cline with $v > 0$.
In Figure 1(a), we show the exact solution (solid line); the
singular-perturbation or $n=0$ solution (dashed line); 
the $n=1$ solution (dotted line); and the $n=2$ solution 
(dot-dashed line). The $n=3$ perturbative solution (dot-dot-dashed 
line) is already numerically indistinguishable from the exact solution 
on the scale of the figure -- we
will quantify the error shortly. We have numerically confirmed that
the perturbation series in Eq.~(\ref{psol})
is strongly convergent, and the inclusion of
higher-order terms ($n > 3$) does not change the solution appreciably.

Let us next examine the time-dependence of the front velocity.
Recall that the $n=0$ solution did
not exhibit the correct approach to the asymptotic velocity $v=2$.
Figure 1(b) plots $v(t)$ vs.\ $t^{-1}$ for the solutions depicted in
Figure 1(a). Again, we see that the $n=3$ result is almost
coincident with the exact result.

Figures 2(a)-(b) study the evolution of the $d=1$ Fisher equation from
a random initial condition, consisting of 10 randomly-distributed seeds of
random height (between 0 and 0.1). In Figures 2(a)-(b), we focus on
the collision and merger of two fronts, showing the exact solution,
and the solutions for $n=0,1,3$. As in Figure 1(a), the $n=3$ solution is
numerically indistinguishable from the exact solution. 

To quantify the
error involved in our approximations, we compute the ``distance'' between
the exact solution $\psi_{\mathrm{e}} (x,t)$ and an approximate solution
$\psi_{\mathrm{a}} (x,t)$  as follows:
\begin{equation}
\label{dist} D(t) = {1 \over L} \int\limits^{L/2}_{-L/2} \rmd x\;
|\psi_{\mathrm{e}} (x,t) - \psi_{\mathrm{a}} (x,t)|,
\end{equation}
where $L$ is the lattice length. Figure 2(c) plots $D(t)$ vs. $t$ on
a semi-logarithmic scale for
the different solutions depicted in Figures 2(a)-(b). We obtain $D(t)$
by averaging over 25 independent initial conditions constructed
as the random superposition of seeds, Gaussians, sine-cosine
functions, etc. The maximum error for the $n=3$ solution is three orders of
magnitude smaller than that for the $n=0$ (singular-perturbation) 
solution. This quantifies the rapid convergence of our perturbative 
hierarchy. Of course, the error asymptotically approaches zero for
all solutions as $\psi \rightarrow \psi^* = 1$ everywhere.

Next, we consider results for the $d=2$ Fisher equation. Figure 3(a)
shows an evolution snapshot obtained from a seed initial condition
$\psi (x,y,0) = 0.05 \delta (x) \delta (y)$.  The dark circular region
refers to the exact solution, and denotes points where $\psi \geq
0.5$. The solid line refers to the front position for the $n=0$
solution, and is defined by points where $\psi = 0.5$. Figure 3(b)
shows the corresponding variation of the order parameter along a
horizontal cross-section of the lattice.  As in the $d=1$ case, the
singular-perturbation front is appreciably sharper than the exact
front.  For clarity, we do not show the higher-order perturbation
results -- as before, the $n=3$ result is indistinguishable from
the exact solution. We focus on the front position so as to clarify
the convergence of the perturbative solution.  Figure 3(c) plots the
front velocity $v(t)$ vs. $t^{-1}$ for the solutions depicted in
Figure 3(b).

Finally, Figures 4(a)-(b) study the evolution from a random initial
condition for the $d=2$ Fisher equation. The initial condition consists
of 25 randomly-distributed seeds with amplitudes between 0 and 0.1.
Figure 4(a) shows an evolution snapshot at $t=12$. (For clarity, we
only show a $500^2$ corner of the $1000^2$ lattice.) The coding is the
same as that for Figure 3(a). Figure 4(b) shows the variation of the
order parameter along a horizontal cross-section of the snapshot shown
partially in Figure 4(a). We
present results for the exact solution, and approximate solutions with
$n=0,1,3$. The approximation error is quantified in Figure 4(c),
where we plot $D(t)$ vs. $t$ on a semi-logarithmic scale. 
The error is obtained as the $d=2$
generalization of the quantity defined in Eq.~(\ref{dist}). 
As in the $d=1$ case, the error was obtained as an average over 
25 independent initial conditions of different types.

\section{Summary and Discussion}
\label{Summary and Discussion}
Let us conclude this paper with a summary and discussion of the results
presented here. We have studied two important examples of reaction-diffusion
systems, viz., the Fisher equation and the time-dependent Ginzburg-Landau
(TDGL) equation. We are interested in the linearization of these 
and other reaction-diffusion equations, i.e., conversion of the
nonlinear problem to a linear problem.
In both cases, we find that the singular-perturbation
solution is a good starting point for a perturbative expansion.
This expansion transforms the problem of solution of the nonlinear
partial differential equation to the solution of a hierarchy of linear 
partial differential equations.
Our numerical studies demonstrate that this hierarchy
rapidly converges to the exact solution of the relevant equation.
However, we should stress that our primary interests in this paper
are methodological rather than operational.

The techniques developed here are of general applicability to a wide
range of reaction-diffusion equations. Of course, any approximate solution
to the initial-value problem for a given equation
is a good starting point for a perturbative
expansion. However, we find that the singular-perturbation solution
appears to be particularly convenient in that it yields an extremely
accurate solution within a few steps of the perturbation expansion.

\section*{Acknowledgements}

SP is grateful to H.W. Diehl for his kind invitation to visit Essen,
where this work was initiated. We are  grateful to A.J. Bray, R.C. Desai,
H.W. Diehl, K.R. Elder and Y. Shiwa for many useful discussions.
KJW gratefully acknowledges financial support from the Deutsche
Forschungsgemeinschaft under grant Wi-1932/1-1, and additional support
through NSF-grant PHY99-07949.

\begin{figure}
\centerline{\fig{0.8\textwidth}{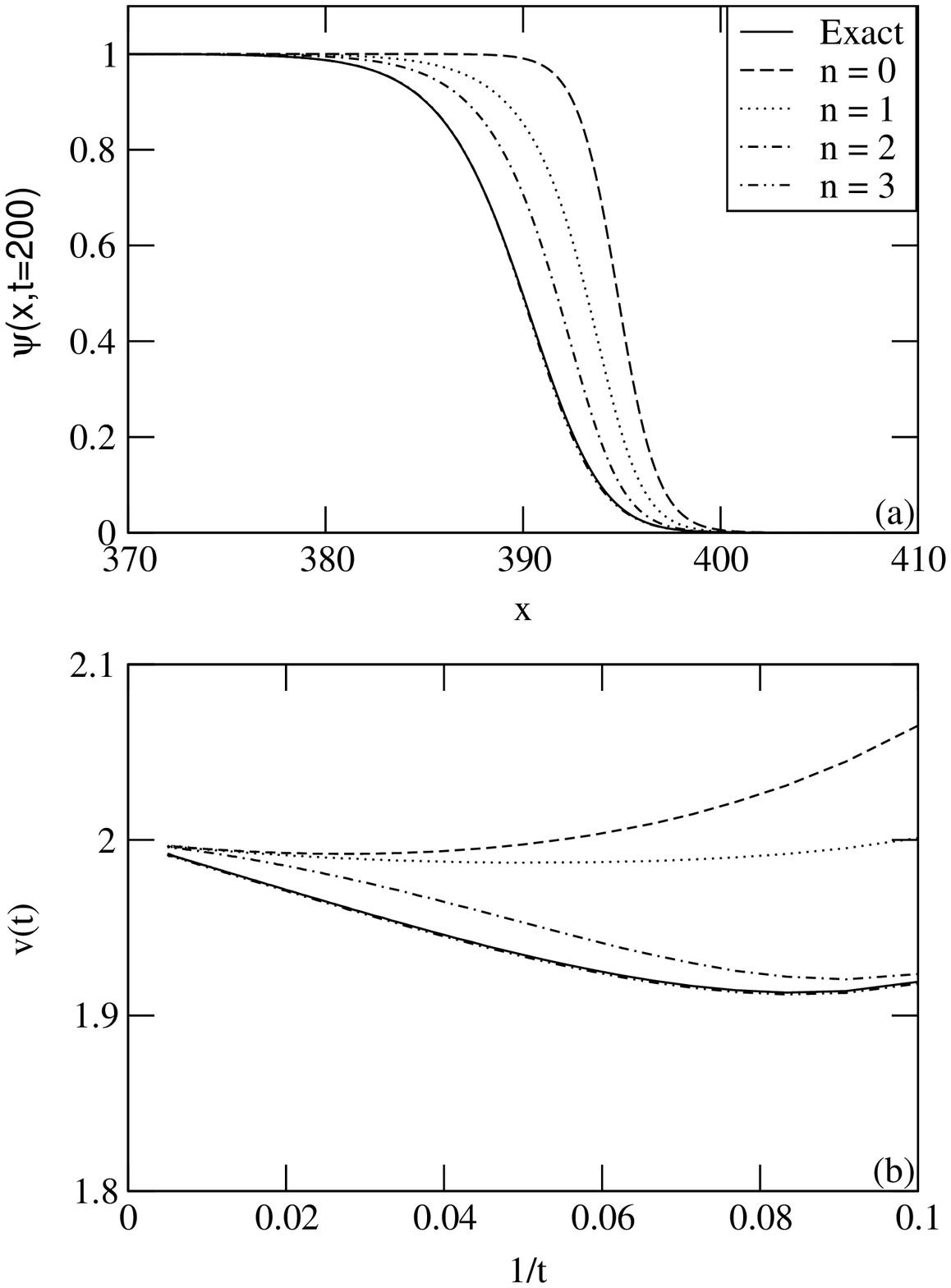}}
\caption{ (a) Evolution snapshot
of the $d=1$ Fisher equation for a
seed initial condition $\psi (x,0)=0.05 \delta (x)$. We plot the order
parameter $\psi (x,t=200)$ vs. $x$ for the front moving with velocity
$v > 0$. Details of our simulation are
provided in the text. We present results for the exact solution
(solid line); the singular-perturbation or $n=0$ solution (dashed line);
the $n=1$ solution (dotted line); the $n=2$ solution (dot-dashed line); 
and the $n=3$ solution (dot-dot-dashed line). 
(b) Plot of front velocity $v(t)$ vs.\ $t^{-1}$ for the solutions
depicted in (a). The line-type usage is the same as that in (a).}
\label{fig1}
\end{figure}

\begin{figure}
\centerline{\fig{0.8\textwidth}{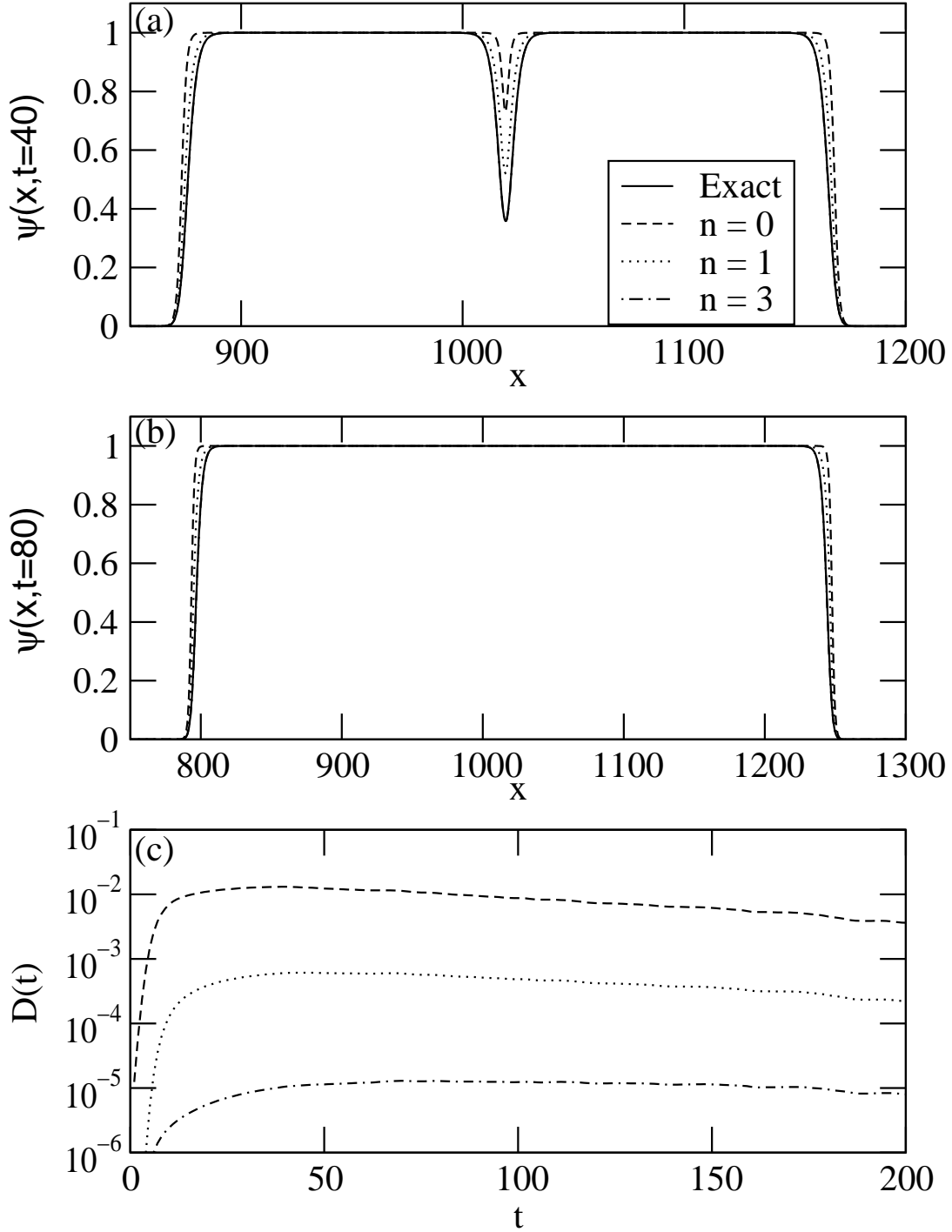}}
\caption{(a) Evolution snapshot
of the $d=1$ Fisher equation for a random
initial condition, consisting of 10 randomly-distributed seeds of amplitude
between 0 and 0.1. We plot the order parameter $\psi (x,t)$ vs. $x$ at $t=40$,
focusing on a front-front collision. The results shown are analogous to 
those in Figure 1(a), except we do not show the case $n=2$. 
(b) Analogous to (a), but at the later time $t=80$. 
(c) Plot of $D(t)$ vs. $t$ on a semi-logarithmic scale, 
where $D(t)$ is the ``distance''
(defined in Eq.~(\ref{dist})) between the exact solution and an approximate
solution. The error $D(t)$ is obtained as an average over 25
independent initial conditions of various types. We show results for the
approximate solutions with $n=0,1,3$, using the same line-types
as in (a)-(b).}
\label{fig2}
\end{figure}
\begin{figure}
\centerline{\fig{0.8\textwidth}{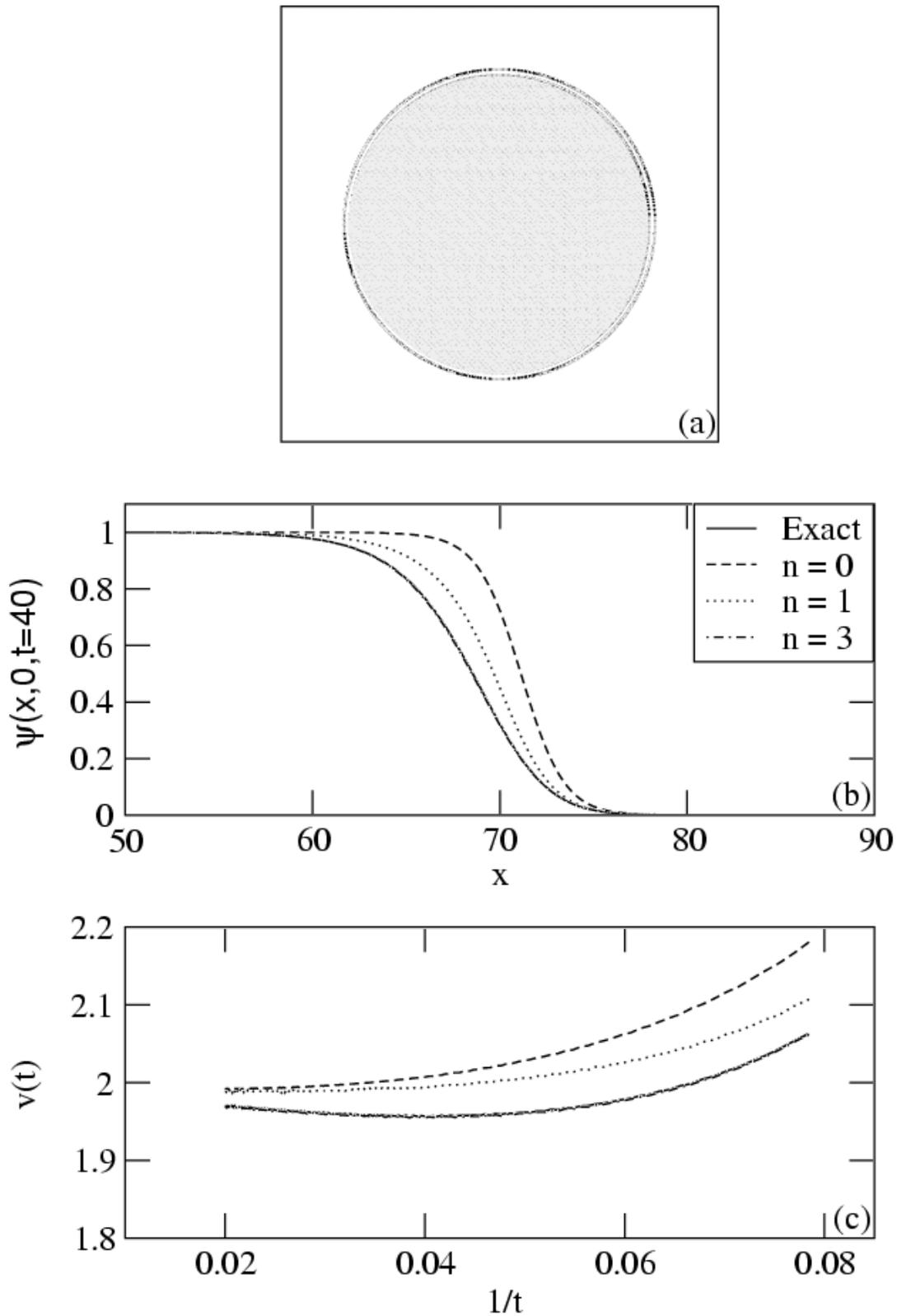}}
\caption{(a) Evolution snapshot
of the $d=2$ Fisher equation from a seed
initial condition $\psi (x,y,0) = 0.05 \delta (x) \delta (y)$. The dark
region corresponds to the exact solution at $t=40$, and denotes points
where $\psi \geq 0.5$. The solid line refers to the
singular-perturbation ($n=0$) solution at $t=40$, and denotes points where
$\psi = 0.5$. 
(b) Variation of the order parameter along a horizontal cross-section
for the evolution depicted in (a). We plot $\psi (x,y,t=40)$ vs. $x$ for
$y=0$, and focus on the front solution. The results presented are analogous
to those in Figure 2(a). 
(c) Plot of front velocity $v(t)$ vs. $t^{-1}$ for the solutions
depicted in (b).}
\label{fig3}
\end{figure}
\begin{figure}
\centerline{\fig{0.8\textwidth}{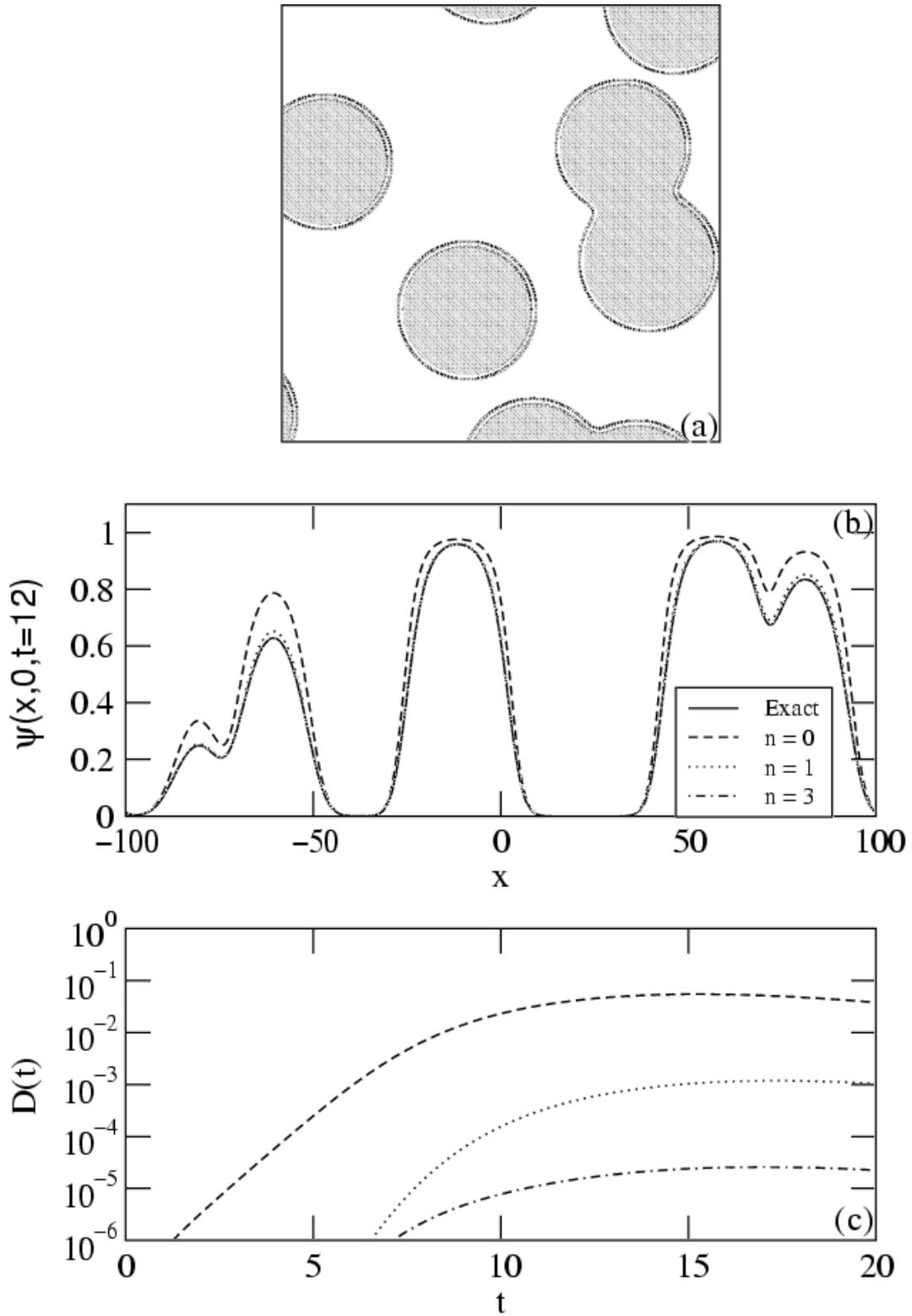}}
\caption{ (a) Evolution snapshot of the $d=2$ Fisher equation 
from a random initial condition, consisting of 25 randomly-distributed 
seeds with amplitude
between 0 and 0.1. The dark regions denote the exact solution at
$t=12$, and the solid line denotes the corresponding singular-perturbation
solution. For clarity, we only show a $500^2$ corner of the $1000^2$
lattice. 
(b) Variation of the order parameter along a horizontal cross-section of the
snapshot partly shown in (a). We plot $\psi (x,y=0,t=12)$ vs. $x$ for 
the entire range of $x$-values. The line-type usage is the same as earlier.
(c) Analogous to Figure 2(c), but for the $d=2$ Fisher equation.}
\label{fig4}
\end{figure}

\end{document}